# Meron-Like Topological Spin Defects in Monolayer CrCl$_3$

Xiaobo Lu[1], Ruixiang Fei[1], Linghan Zhu[1], and Li Yang[1,2] *

[1]Department of Physics, Washington University in St. Louis, St. Louis, MO 63130, USA.

[2]Institute of Materials Science and Engineering, Washington University in St. Louis, St. Louis, MO 63130, USA.

*Email: lyang@physics.wustl.edu

Abstract

Noncollinear spin textures in low-dimensional magnetic systems have been studied for decades because of their extraordinary properties and promising applications derived from the chirality and topological nature. However, material realizations of topological spin states are still limited. Employing first-principles and Monte Carlo simulations, we propose that monolayer chromium trichloride (CrCl$_3$) can be a promising candidate for observing the vortex/antivortex type of topological defects, so-called merons. The numbers of vortices and antivortices are found to be the same, maintaining an overall integer topological unit. By perturbing with external magnetic fields, we show the robustness of these meron pairs and reveal a rich phase space to tune the hybridization between the ferromagnetic order and meron-like defects. The signatures of topological excitations under external magnetic field also provide crucial information for experimental justifications. Our study predicts that two-dimensional magnets with weak spin-orbit coupling can be a promising family for realizing meron-like spin textures.

## Introduction

The quest for low-dimensional ($d \leq 2$) magnetic materials and their unique topological spin textures can be traced back to the last century. Particularly, for two-dimensional (2D) magnetic systems, different ways to break the continuous rotational spin symmetry may induce different spontaneous magnetic orders by violating the Mermin-Wagner theorem.[1–9] For example, with the presence of a specific easy axis in 2D structures, the long-range ferromagnetic (FM) order can sustain at finite temperature via opening a magnon gap to resist thermal agitations.[3,5–10] This is evidenced by recent realizations of Ising-like 2D magnets, such as monolayer (ML) FM insulators $CrI_3$ and $Cr_2Ge_2Te_6$, and multilayer magnetic topological insulator $MnBi_2Te_4$, etc.[6,7,11–14], which have ignited tremendous research interests to date.

Unlike the easy-axis anisotropy, an easy-plane anisotropy yields a residual SO(2) symmetry that usually prohibits the formation of a spontaneous long-range magnetic order at finite temperature.[1,2,4] On the other hand, such a residual in-plane symmetry provides ample space for realizing the quasi-ordered states that evolve with various topological defects. A known example is the topological vortex/antivortex pairs with an algebraically decaying correlation, which can be described by the Berezinskii-Kosterlitz-Thouless (BKT) theory based on the 2D XY model.[4,15] However, spins of realistic materials own a three-dimensional (3D) degree of freedom, making them different from the ideal XY model, in which the spin is confined within the 2D easy plane. Such an extra degree of freedom gives hope to many nontrivial topological spin states, such as skyrmions, magnetic bubbles, and merons.[5,16–21]

These noncollinear spin textures can be induced by many mechanisms, such as antisymmetric exchange interactions (Dzyaloshinskii-Moriya (DM)), magnetic anisotropy, and magnetic dipolar couplings.[5] Particularly, isotropic easy-plane magnetized 2D materials can host magnetic vortices and antivortices. This type of topological defects were predicted to have curling magnetizations lying within the easy plane but pointing out of the plane around the core regions of vortices/antivortices.[5] In condensed matter physics, such a topological structure is referred to as a meron, which corresponds to one half of a skyrmion and is stabilized via pairs.[22–24] Merons have already been observed in various coupled magnetic discs and magnetic interfaces.[18,23–27] For example, meron textures have been observed in thin plates of chiral-lattice magnet $Co_8Zn_9Mn_3$ and

α-Fe$_2$O$_3$/Co heterostructure film at room temperature.[25,27] Recently, there was also a report about creating and stabilizing meron pairs in a continuous Py film, which inherits the in-plane anisotropy by local vortex imprinting from a Co disk.[28] However, besides magnetic discs and interfaces, a pristine single-atomic thin material that exhibits such topological defects is still absent. More importantly, the understanding of these topological defects, their interactions, and the profound relations with fundamental quantum phenomena, such as superfluidity and superconductivity, are yet unclear and forging ahead.[29,30]

## Results

In this work, we find that, because of the weak spin-orbit coupling (SOC), the magnetic dipolar interaction induced magnetic shape anisotropy (MSA) can overcome the magneto-crystalline anisotropy (MCA) to evince an easy-plane, isotropic magnetic polarization in ML CrCl$_3$. By employing an anisotropic Heisenberg model with magnetic dipole-dipole (D-D) interactions and exchange interactions extracted from first-principles simulations, we predict the existence of meron-like topological defects in ML CrCl$_3$. Moreover, beyond pairs of vortecies and antivortecies, higher-order states involved with more than two merons are also observed in our simulations, forming complex topological excitations. Finally, to guide experiments, we show that merons are robust against external fields, and the rich hybridization between merons and the FM order can be tuned via external magnetic fields.

**Easy-plane anisotropy in ML CrCl$_3$.** The easy-plane anisotropy is crucial for the formation of topological defects.[5,31] Usually, the anisotropy is described by the magnetic anisotropy energy (MAE), which characterizes the dependence of energy on the orientation of magnetization. There are two origins of MAE owing to the relativistic effect. The first is MCA that is mainly determined by the SOC. The second is the shape anisotropy related to MSA that is originated from the Breit modification of the relativistic two-electron energy. MSA is ascribed to the classical magnetic D-D interactions:[32,33]

$$E_{\text{D-D}} = -\frac{1}{2}\frac{\mu_0}{4\pi}\sum_{i \neq j}\frac{1}{r_{ij}^3}\left[\mathbf{m}_i \cdot \mathbf{m}_j - 3\frac{(\mathbf{r}_{ij} \cdot \mathbf{m}_i)(\mathbf{r}_{ij} \cdot \mathbf{m}_j)}{r_{ij}^2}\right], \qquad (1)$$

where $\mu_0$ is the vacuum permeability, and $\mathbf{m}_i$, $\mathbf{m}_j$ are the local magnetic moments with a spatial separation of $\mathbf{r}_{ij}$. Such a D-D interaction usually favors magnetization along the elongated direction of materials. For thin films, this provides an origin of in-plane anisotropy. [5,32] Typically, the shape anisotropy can be added as a posterior term after the first-principles electronic calculations. [33] In most materials, the magnetic D-D interaction is small compared with the MCA interaction.[34] However, it may play an important role in weak-MCA magnets with negligible SOC. This leads us to study ML CrCl$_3$, which is known for its weak SOC. [33] 2D CrCl$_3$ has been recently fabricated, [35–37] and the XY physics was discussed in a recent experiment on ML CrCl$_3$ [38].

According to Eq. (1), the MSA energy is calculated by $E_{\text{MSA}} = E_{\parallel}^{\text{D-D}} - E_{\perp}^{\text{D-D}}$ with the magnetization rotating from the in-plane direction ($\parallel$) to the out-of-plane direction ($\perp$). The MCA energy is calculated by $E_{\text{MCA}} = E_{\parallel}^{\text{SOC}} - E_{\perp}^{\text{SOC}}$, which is obtained from first-principles density functional theory (DFT)+U calculations. The calculation setups and convergence tests are included in the Method section. The schematic plot of the FM anisotropy energy (taking the out-of-plane direction as a reference) is presented in Figure 1 (a). Because of the competition between MCA and MSA, the overall MAE of ML CrCl$_3$ is about -34 μeV (the negative sign means a preferred in-plane polarization). Thus, the zero-temperature ground-state magnetization in ML CrCl$_3$ is in-plane (Figure 1 (c)). Besides, the in-plane MAE is nearly perfectly isotropic, and the variation is less than 0.1 μeV. In this work, we choose the Hubbard U and Hund exchange J as 2.7eV and 0.7eV, respectively, which were used in published literatures and provided good agreements with available measurements.[3,39–41] Moreover, we have checked other choices of U and J within a reasonable range and confirmed that they do not qualitatively affect the preferred in-plane MAE and subsequent topological spin textures of ML CrCl$_3$ (see the Supplementary Note 1 and Supplementary Table 1). These results also agree with recent calculations and measurements.[33,35–37]

For the purpose of comparison, we have done calculations for ML CrI$_3$, which has a large SOC. Its MAE is around 300 μeV, resulting in a known Ising-like out-of-plane magnetic order.[3,6,42,43] The corresponding anisotropy energy and preferred zero-temperature out-of-plane magnetization are schematically plotted in Figures 1 (b) and (d). The calculated anisotropy energies and magnetic coupling constants are listed in Table I for these widely studied chromium trihalides, and ML CrCl$_3$ is the most promising structure to realize the in-plane ground-state polarization. Moreover, the relationship between the MCA energy and SOC strength is discussed in the Supplementary Note 2 to show that SOC is the main mechanism in deciding the MCA energy.

**Low-temperature magnetic phase.** To explore the magnetic phase at low temperature, we have employed a classical Monte Carlo (MC) simulation based on an anisotropic Heisenberg model with parameters from first-principles simulations. This Heisenberg Hamiltonian is essentially an approximation to the many-body electronic Hamiltonian in a localized basis, which only contains the spin degree of freedom.[44] This method provides good agreements of the Curie temperature ($T_c$) for ML CrI$_3$ and CrBr$_3$ with experiments.[3,9,41,44,45] More specifically, we add the magnetic D-D interaction ($E_{D-D}$) from Eq. (1):

$$\mathcal{H} = \sum_i A(S_i^z)^2 + \sum_{<i,j>} \frac{1}{2}(\lambda_1 S_i^z S_j^z + J_1 \mathbf{S}_i \cdot \mathbf{S}_j) + \sum_{\ll i,j \gg} \frac{1}{2}(\lambda_2 S_i^z S_j^z + J_2 \mathbf{S}_i \cdot \mathbf{S}_j) + E_{D-D}$$
$$+ \sum_i \mathbf{B} \cdot \mathbf{S}_i, \quad (2)$$

where $A$ describes the easy-axis single-ion anisotropy, $\lambda_{1,2}$ and $J_{1,2}$ represent the anisotropic and isotropic exchange couplings up to the next nearest neighbors, and $\mathbf{B}$ is the external magnetic field. The details of extracting those coefficients from first-principles simulations are presented in the Method section.

We begin with the intrinsic ML CrCl$_3$ without an external field. The evolution of the averaged magnetic moment with temperature for ML CrCl$_3$ is presented in Figure 2 (a). We find that the out-of-plane magnetization (the blue-dotted line $M_z$) is almost completely quenched. Surprisingly, the in-plane magnetization (the red diamond $M_{in}$) with large fluctuations emerges even after the result is averaged over 25 ensembles. In contrast to CrCl$_3$, as shown in Figure 2 (b), ML CrI$_3$ exhibits a normal Ising-like out-of-plane magnetism with negligible fluctuations, and the in-plane magnetization is always zero. A sharp phase transition is observed around 42 K for ML CrI$_3$, which is close to the measured Curie temperature of 45 K.[6]

The inset of Figure 2 (a) further addresses the substantial fluctuations of the in-plane magnetization by comparing the statistically averaged in-plane magnetic moment with a single-round simulation. Importantly, for a single-round simulation, there is no obvious trend of the evolution of the in-plane polarization even between adjacent temperatures. Such a remarkable randomness strongly indicates the existence of a weak polarized state or a quasi-ordered phase, *e.g.*, the vortex/antivortex states described by BKT physics. [4,15,46] The detailed comparison of averaged magnetization for ML $CrCl_3$ and $CrI_3$ with and without D-D interactions can be found in the Supplementary Note 3.

We have also checked the potential existence of quasi-order in monolayer $CrCl_3$ by analyzing the spin-spin correlation functions, which is an effective way to identify orders. [15,47] In Figures 2(c), the spin correlation of ML $CrCl_3$ exhibits a quasi-long-range algebraic decay to zero at a low temperature (4K) and an exponential decay to zero at a high-temperature (12K) that is within the paramagnetic (PM) phase. These behaviors indicate that 1) there is no long-range order in ML $CrCl_3$; 2) quasi orders may exist at low temperature because of the algebraically decaying correlation function. On the other hand, ML $CrI_3$ exhibits a normal FM behavior in Figure 2(d): the spin correlation exponentially decays to a non-zero value at 40K because of the long-range FM order while it exponentially decays to zero at 48K because of the PM phase.

In the following, by further analyzing the real-space arrangement of local magnetic moments as obtained through MC simulations, we confirm the topological vortex/antivortex type of defects within a 3D spin space. Figures 3 (a), (b), and (c) present the real-space magnetic moments at different temperatures. At low temperatures (Figures 3 (a) and (b)), we can identify the vortex/antivortex type of topological spin defects, which are marked by the ovals. At higher temperature (Figure 3 (c)), the spin structures disappear. Importantly, the numbers of vortex and antivortex defects are always the same, maintaining an overall integer topological unit. The chirality of meron pairs is indicated by the blue (vortex type) or red (antivortex type) part of the ovals. Additionally, we have plotted the corresponding relative phase maps, which are used in experiments to identify spin textures.[48,49] In Figures 3 (d), (e) and (f), the phase maps clarify the chirality of topological defects at core regions.

At low temperature (0.25K), we find that these vortex (antivortex) type spin defects can be connected with neighbor alternative antivortex (vortex) type defects via a spin flux closure as indicated by the dashed line in Figure 3 (d). The enlarged Figure 3 (g) provides a clear microstructure diagram of those topological defects. Interestingly, unlike the strict XY model where spin rotator is constrained within the 2D XY plane, the topological structures in ML $CrCl_3$ contain 3D information, and this 3D spin effect is particularly significant around the core regions of the topological structures. As seen from the side view in Figure 3 (h), the magnetic moments around cores point out of the plane and form alterable spin hills.[26,28,49–51] Compared with the vortex and antivortex pairs in the XY-model, whose free energy mainly comes from the core regions[4,15], the benefit of forming such 3D spin texture is manifest by avoiding large swirl angles around the cores of defects and, hence, lowering the system energy. Meanwhile, our simulation does not display any correlation between the magnetic moment swirling and the direction of the central spin hill. This agrees with the experimental observations that all four possible up/down combinations of paired meron-core polarities show up with nearly equal probabilities.[28]

Such topological defects have been referred to as merons, which have a $\pm 1/2$ skyrmion charge [5,23,52] and are similar to experimental observations.[18,23–26] This is a material analogy to a solution of the Yang-Mills theory, in which merons as described in the context of quark confinement can only exist in pairs owing to the one half of topological charge carried.[22,23,52,53] This character agrees with the equal numbers of vortex/antivortex defects observed in our simulations. Particularly, because of the pairing character of these topological spin defects, it is expected that doped free carriers could be intrinsically spin-polarized, and the transport of charged meron pairs might be dispersionless at low temperature. [54–56] Although those dynamic behaviors and quantitative interactions between merons are beyond the scope of this study, they would be interesting topics for future studies.[57–59]

Beyond meron pairs, our MC simulations reveal more complicated spin textures, such as the higher-order states involved more than two merons. As indicated in Figures 3 (a) and (d), some pairs of vortex/antivortex type merons are found to be anti-parallelly aligned with each other at low temperature, forming "quadrupole-like" topological excitations. [57,60] The presence of these hierarchical excitations indicates that, although the classical MC treatment does not take into account the quantum nature of spins, it does include the correlations of magnetic moments from the Heisenberg model. [9] (see Supplementary Fig. 4 for more examples of coupled meron pairs)

As temperature increases, those higher-order excitations are firstly annihilated by thermal fluctuations as indicated by the snapshot at 3.25K in Figures 3 (b) and (e). On the other hand, the relatively robust pairing between vortex/antivortex type merons still persists, although thermal energy blurs the topological structures and makes them harder to be recognized in Figure 3 (e). Finally, at the higher temperature (~ 12K), the paired merons totally melt, as shown in Figures 3 (c) and (f). The system becomes a disordered, PM phase with an exponentially decaying spin-spin correlation, which agrees with Figure 2 (c).

There are a few important mechanisms that are crucial for understanding the simulation results. First, we clarify the role of D-D interactions in generating meron-like spin textures. It is known that D-D interactions complicate system behaviors due to its nonlocal character and weakened singularities of the longitudinal mode of spin waves.[61–63] For 3D spins in ML $CrCl_3$, D-D interactions mainly impact the MAE and introduce the preferred in-plane ground-state polarization. This is evidenced by the similar averaged magnetic moments calculated with different truncations of D-D interactions in Figure 2 (e). In the above simulations, we truncate the D-D interaction with a cutoff radius around 31 Å, as adopted in previous studies. [50,64] When the truncation radius increases from 15Å to 52 Å, the statistically averaged value of magnetic moment at a certain temperature ($T$=3K) does not change essentially. Moreover, we find that the residual D-D interaction beyond a 31 Å truncation contributes less than 5% to the total MSA energy. Thus, the overall D-D interaction energy is, at least, an order of magnitude smaller than the energy from the isotropic exchange interaction $J$. As a result, except at extremely low temperature, the residual energy due to the D-D interaction truncation shall not qualitatively affect the spin textures of our studied systems. It is worth mentioning that our simulation cannot give accurate asymptotical behaviors approaching zero temperature due to the exponentially increasing simulation time, but this limit does not affect the main results of this work.

The simulation-size effect also needs to be considered. As Bramwell and Holdsworth discussed, a spontaneous magnetization could show up in a finite-size XY model owing to the suppression of

Goldstone modes that usually destroy ferromagnetism.[63,65] A similar behavior could happen in 3D-spin cases. As shown in Figure 2 (f) at 3K, when the simulation size is not big enough (less than ~400 Å), the averaged magnetic moment of ML CrCl$_3$ remains nearly a constant and does not exhibit large fluctuations, resembling the typical FM order. This also agrees with previous anisotropic Heisenberg-model studies[64,66] on planar phases of 2D lattices, in which only the planar FM type curve was observed due to small simulation sizes. On the other hand, when the simulation size is significantly larger than that of meron pairs (~ 200 Å estimated from Figure 3(a)), quasi orders start to show up because the system is large enough to hold meron-pairs. Particularly, with simulation size increasing, owing to the geometric features and growing density of merons, the averaged magnetization is expected to gradually decrease. This is what we observed in Figure 2 (f): when the system size increases, the averaged in-plane magnetic moment of ML CrCl$_3$ (the deep blue line with red circles) decreases with large fluctuations. In contrast, the magnetic polarization of Ising-like ML CrI$_3$ (the green line with light blue circles) does not depend on the system size, showing a normal FM state. Therefore, we conclude that a finite-size simulation can capture those meron pairs if the system size is significantly larger than that of merons. In other words, larger systems can hold more meron pairs but do not change the existence of the meron-like phase.

Finally, we also check the proximity effect from substrates, such as the widely used hexagonal boron nitride (h-BN). We do not find significant modifications to the meron states when ML CrCl$_3$ is attached to h-BN, as shown in the Supplementary Table 2, Supplementary Fig. 3, and Supplementary Note 4. This verifies the feasibility of experimental realization for our theoretical predictions.

**Response to external field.** It is highly motivated to further investigate the response of the topologically paired merons to an external static magnetic field, which has been widely employed to study magnetic properties of materials.[6,7,35,37] As discussed above, the polarization direction of the local magnetic moments is crucial for determining the existence of the meron phases. As a result, a magnetic field applied along different directions is expected to induce different responses. First, we apply a small in-plane magnetic field (70 Gs), and the results are shown in Figure 4 (a). At low temperature (below 2K), the magnetic moment is close to the saturated value ($3\mu_B$/Cr), and the fluctuation is substantially quenched, indicating a well-defined FM order. This is confirmed by Figures 4 (c) and (g), in which both the real-space magnetic moment distribution and projected magnetic component along the external-field direction show a FM order. This suggests that the applied in-plane magnetic field breaks the in-plane isotropy, forming a preferred polarization direction, which stabilizes the FM order. Interestingly, as temperature goes up, the magnetic polarization decreases, and larger fluctuation accompanies. In the real-space snapshot (Figure 4 (d)), meron pairs emerge. Those thermally excited topological defects are embedded in the aligned spin sea, which is tilted along the direction of the external field. Figures 4 (d) and (h) provide a physics picture analogy of spin distributions in the real space and projected space. The latter can be directly compared with the experimentally observed figure format.[48,49]

We then explore the case where the applied magnetic field is perpendicular to the material plane. If the out-of-plane magnetic field is below a critical value ($B_\perp^{\text{crit}} \approx 0.2\text{T}$), although it is not strong enough to completely govern the preferred magnetization direction, the local magnetic moments are no longer perfectly within the material plane. Consequently, we observe a net averaged magnetization along the out-of-plane direction. In this weak magnetic field situation, the topological meron pairs may be preserved. Take $B_\perp = 0.1\text{T}$ as an example shown in Figure 4 (b). At low temperature, an averaged out-of-plane magnetization shows up with the help of external field (the orange-dotted curve $M_z$). Meanwhile, the in-plane magnetic moment (the red-dotted curve $M_{xy}$) exhibits large fluctuations, indicating the existence of excited merons. This is evidenced by both the real-space and projected component plots of the magnetic moments in Figure 4 (e) and (i). The hot spots in the projected component figure indicate the core regions of merons, which correspond to the previously discussed spin hills at meron cores. When the temperature reaches 10K, the averaged in-plane magnetization approaches zero. On the other hand, the out-of-plane magnetization is preserved, as shown in Figure 4 (f) and (j). This is another type of hybridization between FM and meron states. Moreover, the reason for the abnormal elevation of out-of-plane FM polarization with increasing temperature at around 10K is that the D-D interaction is more sensitive to the arrangements of adjacent magnetic moments than the on-site Zeeman energy provided by the external field.

**Discussion**

In summary, our studies show that the overall isotropic in-plane magnetized 2D material such as ML CrCl$_3$ could provide the opportunity for generating meron-type topological defects without involving any other interactions. This would give hope to realizing topological spin structures in a much broader range of materials with weak SOC, such as CrF$_3$. Meanwhile, these meron-like topological spin defects may shed light on the understanding of the recently observed suppressed magnetic ordering in ML NiPS$_3$, an AFM easy-plane anisotropy material.[67] In addition to the previously studied Josephson junction arrays or superfluid helium systems.[15,29,54], these meron-like pairs may provide an alternative way to study the exotic relations between topological spin textures and quantum critical phenomena, including the superfluidity and superconducting behaviors, in pristine 2D magnetic systems.

## Methods

**Density functional theory (DFT) calculations.** The DFT calculations are performed within the generalized gradient approximation (GGA) using the Perdew-Burke-Ernzerhof (PBE) functional as implemented in the Vienna Ab initio Simulation Package (VASP).[68] A plane-wave basis set with a kinetic energy cutoff of 450eV and a 5x5x1 k-point sampling is adopted for a 2x2x1 supercell to mimic different magnetic configurations for extracting magnetic interaction constants. The k-point grid is 12x12x1 for a 2x2x1 supercell. The vacuum distance is set to be 20 Å between adjacent layers to avoid spurious interactions. The van der Waals (vdW) interaction is included by the DFT-D2 method,[69] and SOC is considered. We choose the Hubbard U = 2.7eV and Hund J = 0.7eV for Cr$^{3+}$ ions. The structure is relaxed until the force converges within 0.01 eV/Å. The optimal lattice constant is 6.01 Å, which is close to the bulk monoclinic structure lattice constant 6.06 Å.[70]

**Determination of the Heisenberg Hamiltonian coefficients.** We obtain the magnetic interaction coefficients of the anisotropic Heisenberg model Hamiltonian (Eq (2)) by calculating the total energies of different magnetic configurations. We consider the FM and Néel antiferromagnetic (AFM) configurations. The corresponding energies for a unit cell are

$$E_{FM/AFM}^{out} = E_0 + (2A \pm 3\lambda_1 \pm 3J_1 + 6\lambda_2 + 6J_2)|\mathbf{S}|^2 \quad (3),$$

$$E_{FM/AFM}^{in} = E_0 + (\pm 3J_1 + 6J_2)|\mathbf{S}|^2 \quad (4),$$

in which two magnetic orientations (in-plane and out-of-plane) are calculated to determine the anisotropic coupling constants. Moreover, we flip the magnetic moment of a $Cr^{3+}$ cation in the 2x2x1 supercell to realize more magnetic configurations (the energies are normalized to one unit cell).

$$E_{flip}^{out} = E_0 + \left(2A + \frac{3}{2}\lambda_1 + \frac{3}{2}J_1 + 3\lambda_2 + 3J_2\right)|\mathbf{S}|^2 \quad (5)$$

$$E_{flip}^{in} = E_0 + \left(\frac{3}{2}J_1 + 3J_2\right)|\mathbf{S}|^2 \quad (6)$$

As a result, we obtain six equations, which are used to solve for the five magnetic interaction coefficients in Eq (1) and the reference energy ($E_0$). The extracted coefficients and corresponding magnetic anisotropy energies for $CrX_3$ (X=Cr, Br, I) are summarized in the Table 1.

**MC simulations.** Based on the Hamiltonian Eq. (2), we perform MC simulations via the Metropolis algorithm on 2D hexagonal lattices with a size of 160x160 unit cells (if not particularly specified), which contain 51200 magnetic moments. The periodic boundary condition is implemented. All magnetic moments are set to along the out-of-plane direction at the initial state to mimic experimental cooling conditions with the help of external field. We run for $5.12 \times 10^9$ MC steps ($10 \times 10^5$ steps per site in average) to ensure that the thermal equilibrium is achieved.

The out-of-plane and in-plane magnetizations are defined as:

$$<m_z> = <\frac{1}{N}\sum_{i=1}^{N} S_i^z> \quad (7),$$

$$<m_{xy}> = <\frac{1}{N}\sqrt{\left(\sum_{i=1}^{N} s_i^x\right)^2 + \left(\sum_{i=1}^{N} s_i^y\right)^2}> \quad (8),$$

where $N$ represents the number of total magnetic moments in the simulated system, and $<X>$ denotes the time average of corresponding components after the system achieves the thermal equilibrium.

In magnetic systems, the static spin-spin correlation function describes the average scalar product of spins at two lattice sites with a fixed distance $|\mathbf{r}_1 - \mathbf{r}_2| = r$. It can be written as:

$$C(r) = <\mathbf{S}_1(\mathbf{r}_1) \cdot \mathbf{S}_2(\mathbf{r}_2)> / |\mathbf{S}|^2 \quad (9)$$

**Data Availability**

The data that support the findings of this study are available from the corresponding author upon reasonable request.

**Acknowledgement**

The work is supported by the Air Force Office of Scientific Research (AFOSR) grant No. FA9550-17-1-0304 and National Science Foundation (NSF) CAREER Grant No. DMR-1455346. The computational resources have been provided by the Stampede of TeraGrid at the Texas Advanced Computing Center (TACC) through XSEDE.

**Author Contributions**

L.Y. supervised the project. X.L performed the DFT and MC simulations and drafted the manuscript. All authors discussed the results and edited the manuscript.

**Competing Interests**

The authors declare no competing interests.

**Figures:**

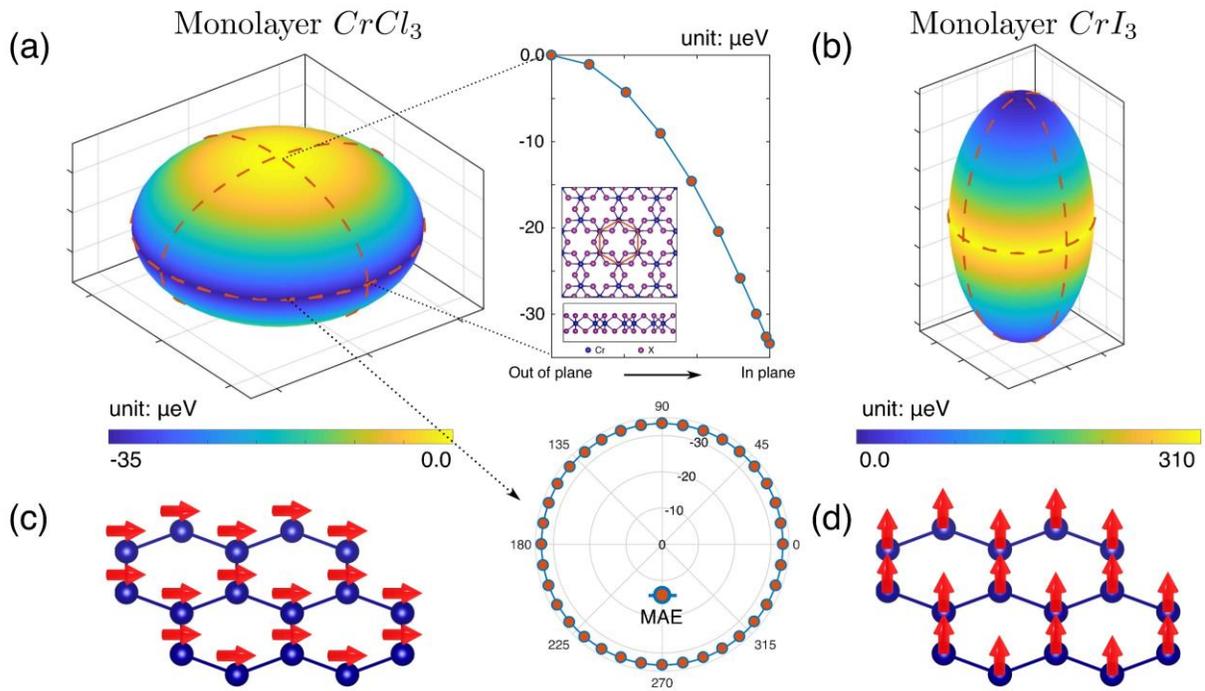

**Figure 1** Magnetic anisotropy energy (MAE) of ML CrCl$_3$ and CrI$_3$. (a-b) MAE map (taking the out-of-plane FM state as a reference) for ML CrCl$_3$ and CrI$_3$, respectively. The figures in the middle panel are extracted from the longitude and latitude from (a) (as indicated by the dashed line), to amplify the variation from the out-of-plane direction to the in-plane direction and the nearly perfect isotropy within the material plane. The inset figure is the atomic structure of chromium trihalides. The red hexagon indicates the honeycomb structure formed by magnetic Cr atoms. (c) and (d) are the schematic plots of the preferred magnetization directions for ML CrCl$_3$ and CrI$_3$, respectively.

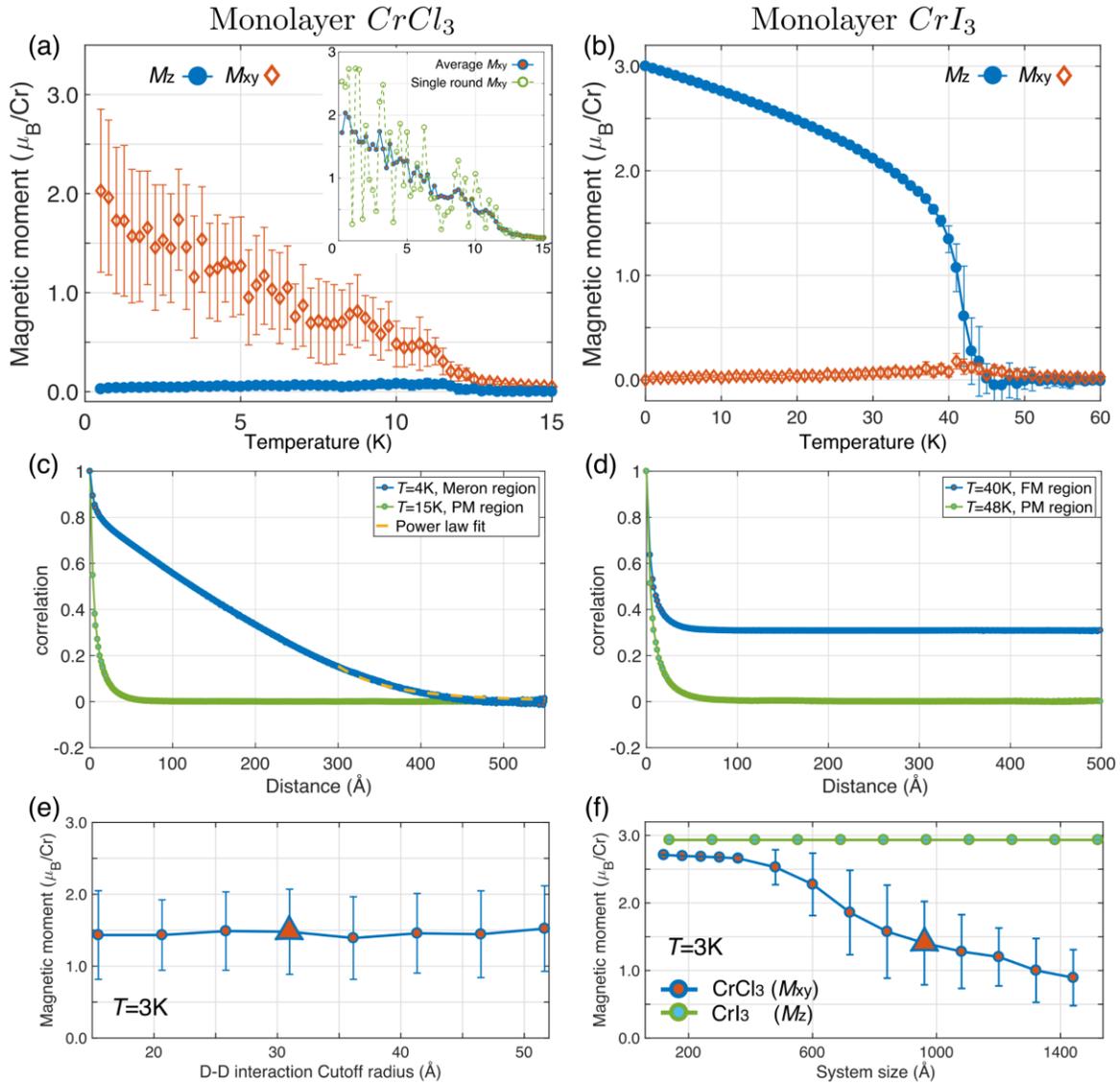

**Figure 2** Monte Carlo simulations of ML CrCl$_3$ and CrI$_3$. (a) and (b) are the MC simulated magnetization versus temperature of ML CrCl$_3$ and CrI$_3$, respectively. The inset of (a) represents the averaged (blue-red line) and single-round (green dotted line) in-plane magnetic polarization of ML CrCl$_3$. (c) and (d) The spin-spin correlation function of ML CrCl$_3$ and CrI$_3$. (e) and (f) The averaged in-plane magnetic polarization at $T$=3K for different D-D interaction truncations and different system sizes. The error bars are the standard deviation.

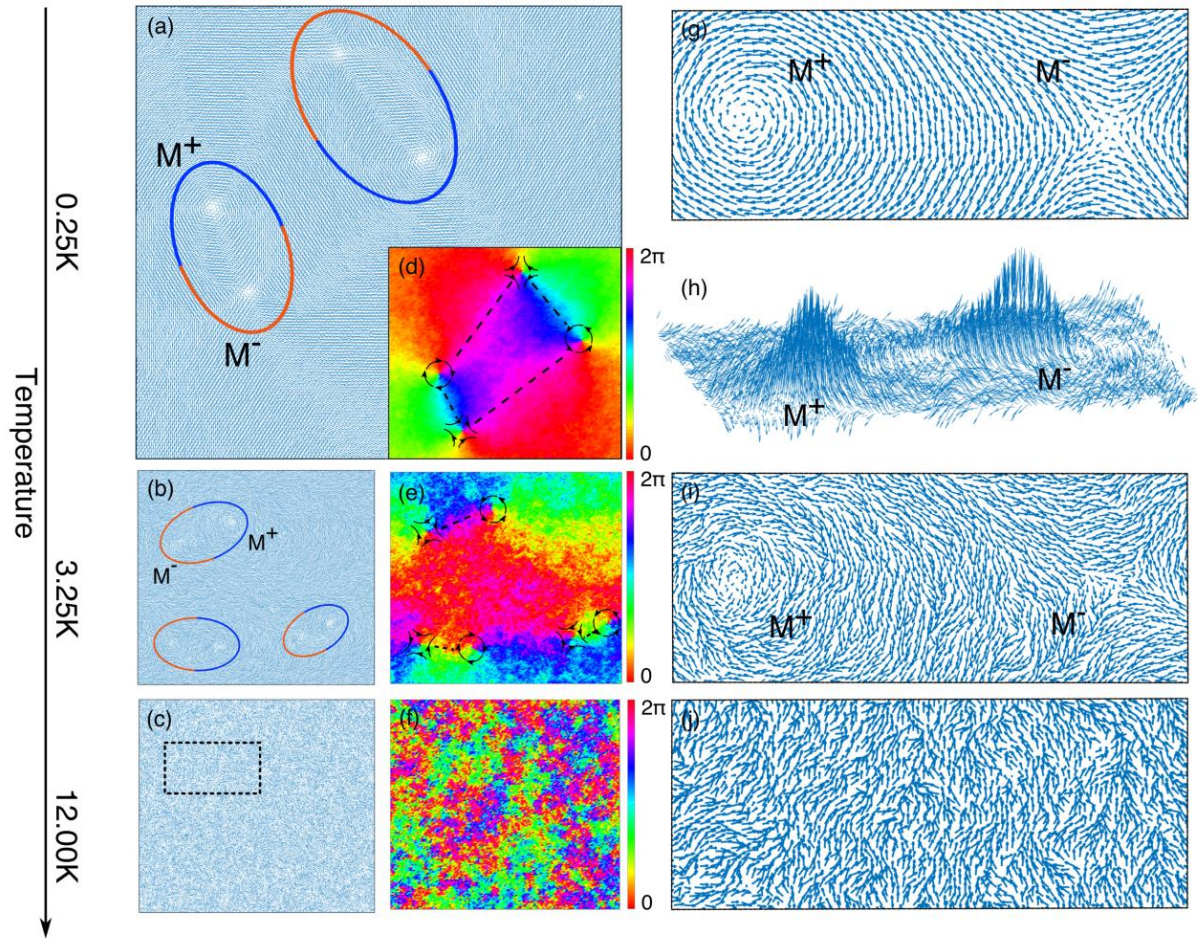

**Figure 3** Snapshots of Monte Carlo simulations of ML $CrCl_3$. (a-c) Top views of the real-space magnetic moments from snapshots of MC simulations under different temperatures. The blue and red part of the ovals indicate the vortex and antivortex type meron defects, respectively. (d-f) are phase maps of in-plane magnetic moment component of (a-c). The schematic spin directions and the connections between defects are drawn to clarify the vortex/antivortex type meron and the higher order excitations. (g), (i), and (j) are enlarged views of the specific defect pairs and the rectangular area from (a), (b), and (c), respectively. (h), Side view of (g) with an enlarged z-axis ratio (10:1).

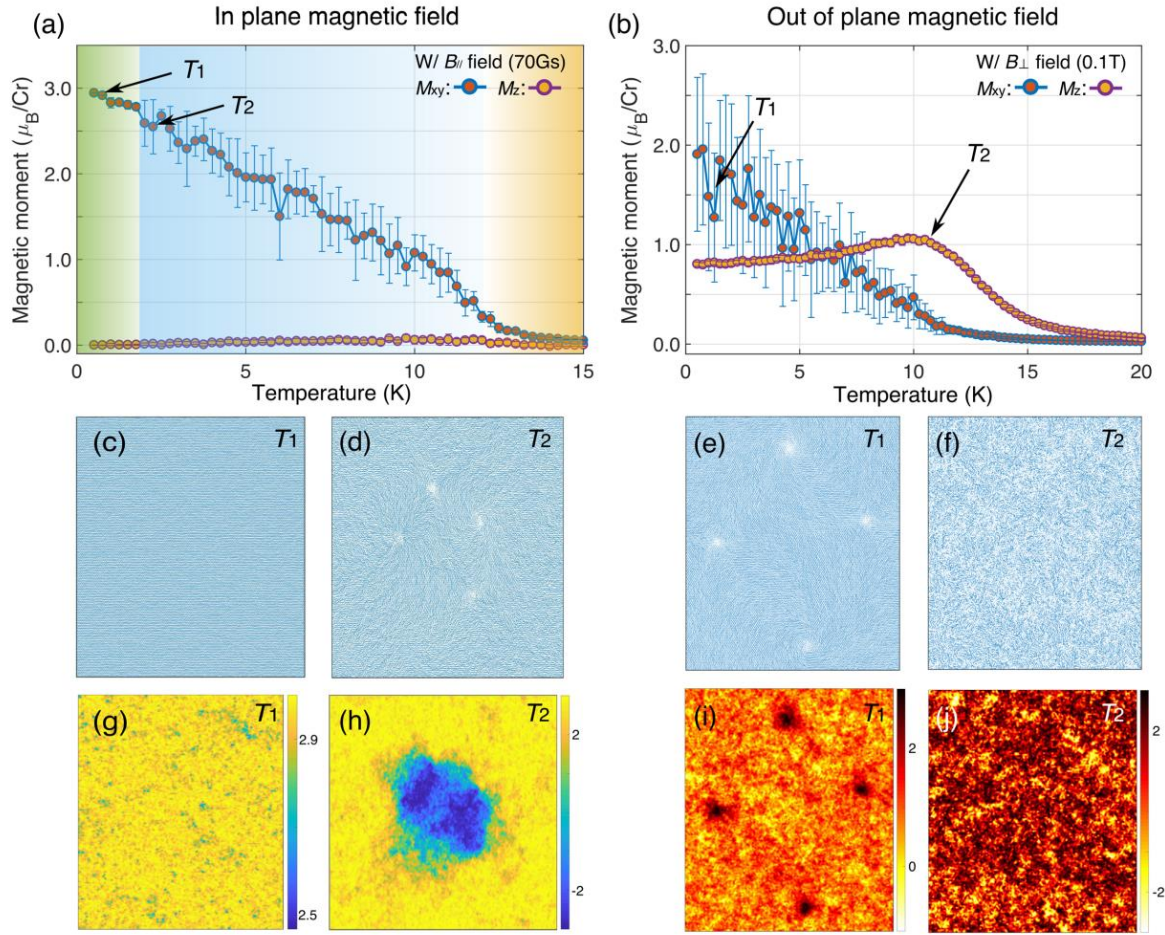

**Figure 4** Response of the spin structures to external magnetic field. (a) and (b) Magnetic polarization versus temperature under parallel and perpendicular external magnetic fields, respectively. The different background colors in (a) indicate different magnetic states. (c-f) Real-space magnetic moments under the marked temperatures in (a) and (b). The lower panels (g-j) are the corresponding projection components of local magnetic moments along external field directions (parallel and perpendicular to the material plane, respectively) from (c-f). The error bars are the standard deviation.

**Tables:**

**Table I** Anisotropy energies and magnetic coupling strengths of ML CrX$_3$ (X=I, Br, Cl).

| Materials | MSA | MCA | MAE | $A$ | $\lambda_1$ | $J_1$ | $\lambda_2$ | $J_2$ |
|---|---|---|---|---|---|---|---|---|
| ML CrI$_3$ | -34 | 349 | 305 | -0.087 | -0.085 | -2.12 | 0.02 | -0.35 |
| ML CrBr$_3$ | -43 | 91 | 48 | -0.020 | -0.016 | -1.35 | -0.001 | -0.153 |
| ML CrCl$_3$ | -54 | 20 | -34 | -0.007 | -0.002 | -0.79 | 0.0004 | -0.071 |

The energy and coupling strength are in μeV and meV, respectively.